\documentstyle[a4,epsf]{article}
\title{\bf Spontaneous CP Violation \\ in Large Extra Dimensions}
\author{{\bf Yutaka Sakamura}\footnote{Email: sakamura@th.phys.titech.ac.jp} \\
{\footnotesize\it Department of Physics, Tokyo Institute of Technology} \\
{\footnotesize\it Oh-okayama, Meguro, Tokyo 152-8551, Japan}}
\date{}
\begin{document}
\maketitle
\begin{abstract}
We show that in the context of large extra dimensions enough CP violation
can be obtained from the spontaneous breaking in a simple non-SUSY model, 
which is usually
considered not to cause the spontaneous CP violation.
We estimate $\epsilon_{K}$ in our scenario to be of order $10^{-3}$ 
consistent with the experimental value.
We also propose a modification to the see-saw mechanism and axion scenario 
to match with our model.
\end{abstract}

\section{Introduction}
String theory indicates the existence of extra dimensions beyond
our usual four dimensional spacetime.
These extra dimensions must be compactified by small radii in order not to
be observed.
However, these radii do not have to be close to the Planck length, but
have only to satisfy the constraint from the gravity experiment, 
i.e., they should be shorter than $cm$ range.
The relation between the fundamental scale $M_{\ast}$ and 
the observed Planck scale $M_{p}$ is given by
\begin{displaymath}
 M_{p}^{2}=M_{\ast}^{n+2}V_{n},
\end{displaymath}
where $V_{n}$ is a volume of the extra space, and
$n$ is the number of the extra dimensions. 
If $M_{\ast}$ is set to TeV scale, $n$ must be greater than one, 
otherwise the gravity would change over the scale of the solar system
from the observed one.

For simplicity, we assume that the shape of the extra space is torus,
in which case $V_{n}=(2\pi)^{n}R_{1}R_{2}\cdots R_{n}$, 
where $R_{i}$ is the radius of the $i$-th extra dimension.
Hence,
\begin{equation}
 M_{p}^{2}=(2\pi)^{n}M_{\ast}^{2+n}R_{1}R_{2}\cdots R_{n}. \label{rlt-pf}
\end{equation}

We can see from Eq.(\ref{rlt-pf}) that $M_{\ast}$ can be lowered from $M_{p}$ 
to TeV scale by taking the radii $R_{i}$ to be large compared to 
the Planck length.
In this way, we can solve the hierarchy problem without supersymmetry (SUSY)
or technicolor \cite{ark-dimo-dv,antoniadis-1}.

We can take the standard model (SM) fields for either bulk fields or 
boundary fields, which are confined to the D-branes or Domain walls.
In the case that the SM fields feel some extra dimensions, 
their compactification scale must be larger than a few hundred GeV 
because corresponding Kaluza-Klein (K.K.) modes have never been 
observed yet \cite{antoniadis-2}.

It is possible to realize the hierarchy among the fermion masses by using 
the ratio of the volume of the region in which the bulk fields spread out 
to that of the boundary fields \cite{yoshioka}.

CP is a very good symmetry, but it is violated in the 
$\mbox{K}$-$\overline{\mbox{K}}$ system by a small amount 
($\epsilon_{K}\simeq 2\times 10^{-3}$).
There are two possibilities for the origin of the CP violation.
One is the explicit CP violation, and the other is the spontaneous CP violation
(SCPV).
In string theory, CP is a gauge symmetry \cite{choi-kaplan,dine-leigh}, 
so it must be spontaneously broken.
If this CP breaking scale is low enough, one finds that CP is violated
spontaneously in an effective field theory at low energy.
We shall consider this case here.

In this paper we shall assume that different fields feel different numbers of
extra dimensions, as discussed in Ref.\cite{yoshioka},
and show in the context of large extra dimensions
enough CP violation can be obtained from the spontaneous breaking 
in a simple non-SUSY model,
which is usually considered not to cause the SCPV.

In Section \ref{model}, we shall explain our model and realize the hierarchy
among the fermion masses.
In Section \ref{cp}, we shall estimate $\epsilon_{K}$ in our model and
show that it is consistent with the observed value.
In Section \ref{neutrino}, the neutrino masses and mixing angles
are derived without a help of intermediate scale, and 
in Section \ref{strong-cp} we shall try to apply the axion scenario to our
context.
Finally, Section \ref{conclusions} contains some conclusions.

\section{Model} \label{model}
\subsection{Our model}
Basically, we shall consider the minimal standard model 
with an additional gauge-singlet scalar field,
but we shall assume that different fields feel different numbers of extra
dimensions.
The relevant interactions are as follows.

\begin{eqnarray}
 {\cal L}_{\mbox{\scriptsize int}}&=&h^{u}_{ij}Q_{i}H^{\dagger}\bar{U}_{j}
 +h^{d}_{ij}Q_{i}H\bar{D}_{j}+h^{e}_{ij}L_{i}H\bar{E}_{j} \nonumber \\
 &&+\hat{y}^{u}_{ij}SQ_{i}H^{\dagger}\bar{U}_{j}+\hat{y}^{d}_{ij}SQ_{i}H
 \bar{D}_{j}+\hat{y}^{e}_{ij}SL_{i}H\bar{E}_{j}+h.c. \nonumber \\
 &&+(\mbox{Higgs sector}), \label{L-int}
\end{eqnarray}
where $Q_{i}$, $\bar{U}_{i}$ and $\bar{D}_{i}$ are the $i$-th generation of 
the left-handed quark doublet, right-handed up-type quark singlet and 
right-handed down-type quark singlet, respectively. 
$L_{i}$ and $\bar{E}_{i}$ are the $i$-th generation of the left-handed lepton 
doublet and right-handed charged lepton singlet.
$H$ and $S$ are the doublet and singlet Higgs fields respectively, 
$h^{x}_{ij}$ ($x=u,d,e$) are
the Yukawa coupling constants and $\hat{y}^{x}_{ij}$ are dimensionful 
coupling constants.

Note that the non-renormalizable terms in the second line of Eq.~(\ref{L-int})
should be considered because the fundamental scale $M_{\ast}$ is relatively
low in our scenario.

We assume the CP-invariance for the Lagrangian, so that all parameters 
in Eq.~(\ref{L-int}) are real.
We will explore the possibility that the CP-invariance is broken spontaneously
at the weak scale due to the complex
vacuum expectation values (VEVs) of the Higgs fields.

The numbers of the extra dimensions that each field can feel are listed in
Table~\ref{nb-exdim}.

\begin{table}
\begin{center}
\begin{tabular}{|c|c||c|c||c|c||c|c||c|c||c|c||c|c|} \hline
 $Q_{1}$ & 2 & $\bar{U}_{1}$ & 2 & $\bar{D}_{1}$ & 1 & $L_{1}$ & 
 2 & $\bar{E}_{1}$ & 2 & $H$ & 0 & $A^{\mu}_{1}$ & 2 \\ \hline
 $Q_{2}$ & 1 & $\bar{U}_{2}$ & 1 & $\bar{D}_{2}$ & 1 & $L_{2}$ & 
 1 & $\bar{E}_{2}$ & 1 & $S$ & 1 & $A^{\mu}_{2}$ & 2 \\ \hline
 $Q_{3}$ & 0 & $\bar{U}_{3}$ & 0 & $\bar{D}_{3}$ & 1 & $L_{3}$ & 
 1 & $\bar{E}_{3}$ & 0 & & & $A^{\mu}_{3}$ & 2 \\ \hline
\end{tabular}
\caption{The number of the extra dimensions that each field can feel.
 $A^{\mu}_{i}(i=1,2,3)$ are gauge fields.} \label{nb-exdim}
\end{center}
\end{table}

The radii of these extra dimensions are supposed to be the same size and
it is denoted by $R_{1}$.

\subsection{Fermion mass hierarchy} \label{ferm-hiechy}
Let us denote $\psi (x,y)$ as a five dimensional bulk field, where 
$y$ represents the coordinate of the fifth dimension compactified by a radius
$R$.
If we Fourier expand 
\begin{displaymath}
 \psi (x,y)=\sum_{m=0}^{\infty}\frac{1}{\sqrt{2\pi R}}\psi_{m}(x)
 e^{i(m/R)y},
\end{displaymath}
then we can regard $\psi_{m}(x)$ as a four dimensional field corresponding to
the $m$-th K.K. mode.

On the other hand, the boundary fields are localized at the four dimensional
wall whose thickness is of order $M_{\ast}^{-1}$,
so a coupling including at least one boundary field is suppressed by 
a factor $\epsilon^{k}$, where $\epsilon\equiv 1/\sqrt{2\pi M_{\ast}R}$ and
$k$ is a number of bulk fields included in the coupling \cite{ark-dim-mch}.
Generalization to our six dimensional case is trivial.

The existence of infinite K.K. modes change the running of the gauge coupling
constants above the compactification scale $R_{1}^{-1}$ to the 
power-law running \cite{taylor}.
So, it seems natural that a new physics like the GUT appears up to one order
above the scale $R_{1}^{-1}$ by considering the runnings of 
the gauge couplings \cite{dienes}.
We shall denote the scale that the new physics appears as $M_{NP}$.

Now Let us assume that coupling constants $h^{x}_{ij}$, $\hat{y}^{x}_{ij}$ are
generated at the scale $M_{NP}$, 
then it seems natural that they are of the form as $\tilde{h}^{x}_{ij}/M_{NP}$
and $\tilde{y}^{x}_{ij}/M_{NP}^{3}$ in the six dimensional bulk spacetime, 
where $\tilde{h}^{x}_{ij}$ and $\tilde{y}^{x}_{ij}$ are 
dimensionless couplings.

In this case the four dimensional couplings $h^{x}_{ij}$ and 
$\hat{y}^{x}_{ij}$ are of the form as
\begin{eqnarray}
 h^{u}_{ij}\simeq \left(\begin{array}{ccc}\epsilon^{4} & \epsilon^{3} &
 \epsilon^{2} \\ \epsilon^{3} & \epsilon^{2} & \epsilon \\
 \epsilon^{2} & \epsilon & 1 \end{array}\right),\;\;\;
 h^{d}_{ij} \simeq \epsilon
 \left(\begin{array}{ccc}\epsilon^{2} & \epsilon^{2} &
 \epsilon^{2} \\ \epsilon & \epsilon & \epsilon \\ 1 & 1 & 1 \end{array}
 \right),\;\;\;
 h^{e}_{ij}\simeq \epsilon\left(\begin{array}{ccc}
 \epsilon^{3} & \epsilon^{2} & \epsilon \\ \epsilon^{2} & \epsilon & 1 \\
 \epsilon^{2} & \epsilon & 1 \end{array}\right) && \nonumber \\
 \hat{y}^{x}_{ij}\simeq\frac{\epsilon M_{\ast}}{M_{NP}^{2}}
 h^{x}_{ij}\hspace{5cm}&&. 
\label{yukawa}
\end{eqnarray}

Here we have assumed $(M_{\ast}/M_{NP})\tilde{h}\simeq 1$
and $\tilde{h}\simeq\tilde{y}$.
Thus, setting $\epsilon\equiv 1/\sqrt{2\pi M_{\ast}R_{1}}$ to be $1/15$,
the desired hierarchy among quark and lepton masses and mixing angles are
obtained.\footnote{Since the energy range of the power-law running is much
smaller than the hierarchy between $M_{\ast}$ and $(2\pi R_{1})^{-1}$, 
the power-law running of the Yukawa couplings does not destroy 
the structure represented by Eq.(\ref{yukawa}).}
For example, if we assume $R_{1}^{-1}\simeq 300$~GeV, we should set 
$M_{\ast}\simeq 10$~TeV.
We shall take these values in the following, and further we shall assume 
$M_{NP}\simeq 3$~TeV.

\section{CP violation} \label{cp}
CP invariance is broken at the weak scale due to the complex VEVs of 
the neutral Higgs fields.
These VEVs are parametrized as
\begin{displaymath}
 \langle H^{0}\rangle=v,\;\;\;
 \langle S\rangle=we^{i\rho},
\end{displaymath}
where $v=174$~GeV, and we have removed the phase of the $\langle H^{0}\rangle$ 
by using the $U(1)_{Y}$ gauge symmetry.
Here note that our scenario does not depend on the Higgs potential, so we
shall assume the potential to have a CP violating minimum.

Then the quark mass matrices at low energy are obtained as
\begin{eqnarray*}
 (M_{u})_{ij}&=&(h^{u}_{ij}+\hat{y}^{u}_{ij}we^{i\rho})v, \\
 (M_{d})_{ij}&=&(h^{d}_{ij}+\hat{y}^{d}_{ij}we^{i\rho})v.
\end{eqnarray*}

We can see from Eq.(\ref{yukawa}) that $M_{u}$ and $M_{d}$ have complex phases
of order 
$\varphi\equiv\epsilon w M_{\ast}/M_{NP}^{2}\simeq 0.03$,
so each element of the CKM matrix also has an $O(\varphi)$ phase.
Here we have assumed $w\simeq 200$~GeV.

Next we expand the neutral Higgs fields around their VEVs as follows.
\begin{eqnarray}
 H^{0}&=&v+\frac{1}{\sqrt{2}}\phi,
 \nonumber \\
 S&=&we^{i\rho}+\frac{1}{\sqrt{2}}e^{i\rho}(X+iY),
\end{eqnarray}
where we have chosen the unitary gauge.
$\phi$ and $X$ are CP-even real scalar fields, and $Y$ is a CP-odd one.

In terms of these fields, renormalizable Yukawa coupling terms below 
the weak scale are
\begin{eqnarray}
 {\cal L}_{\mbox{\scriptsize yukawa}}&=&\frac{m^{u}_{i}}{\sqrt{2}v_{2}}
 q_{i}\bar{u}_{i}\phi
 +\frac{m^{d}_{i}}{\sqrt{2}v_{1}}q_{i}\bar{d}_{i}\phi  
 \nonumber \\
 &&+\frac{1}{\sqrt{2}}y^{u}_{ij}e^{i\rho}q_{i}\bar{u}_{j}(X+iY)
 +\frac{1}{\sqrt{2}}y^{d}_{ij}e^{i\rho}q_{i}\bar{d}_{j}(X+iY)  \nonumber \\
 &&+(\mbox{lepton sector}),
\end{eqnarray}
where $q_{i}$, $u_{i}$ and $d_{i}$ are the mass eigenstates of quarks,
and $m^{x}_{i}$ is the mass eigenvalue corresponding to the state $x_{i}$.
Note that $y^{u}_{ij}\equiv \hat{y}^{u\prime}_{ij}v$ and 
$y^{d}_{ij}\equiv \hat{y}^{d\prime}_{ij}v$ have $O(\varphi)$ phases,
where $\hat{y}^{x\prime}$ is the $\hat{y}^{x}$-matrix in the basis of 
the quark mass eigenstates.

Now we shall estimate $\epsilon_{K}$ in the $\mbox{K}$-$\overline{\mbox{K}}$
system.
The CP violation parameter $\epsilon_{K}$ can be 
expressed as \cite{pomarol,masip-rasin},
\begin{equation}
 |\epsilon_{K}|\simeq\frac{1}{2\sqrt{2}}\frac{{\rm Im}M_{12}}{{\rm Re}M_{12}},
\end{equation}
where $M_{ij}$ is the neutral kaon mass matrix in the 
$\mbox{K}^{0}-\overline{\mbox{K}}^{0}$ basis.

\begin{figure}
 \leavevmode
 \epsfxsize=8cm
 \epsfysize=4cm
 \centerline{\epsfbox{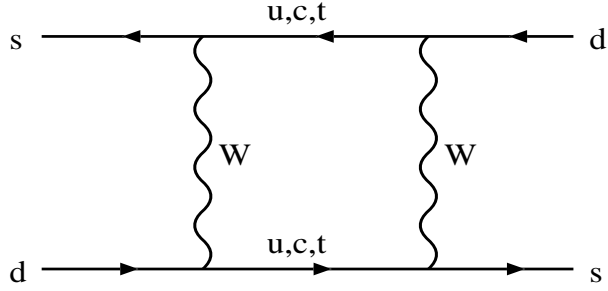}}
 \caption{Dominant contribution to ${\rm Re}M_{12}$.}
 \label{box-dgm}
\end{figure}

The dominant contribution to ${\rm Re}M_{12}$ comes from the standard 
box diagram depicted in Fig.\ref{box-dgm}.

This diagram is estimated as \cite{haber-nir}
\begin{eqnarray}
 M^{\mbox{\scriptsize box}}_{12}&=&\frac{G_{F}}{2\sqrt{2}}
 \frac{\alpha}{4\pi\sin^{2}\theta_{W}}(\cos\theta_{c}\sin\theta_{c})^{2}
 \left(\frac{m_{c}^{2}}{M_{W}^{2}}\right)
 \frac{\langle K^{0}|\bar{d}_{L}\gamma_{\mu}s_{L}\bar{d}_{L}\gamma^{\mu}
 s_{L}|\overline{K^{0}}\rangle}{M_{K}} \nonumber \\
 &\simeq&10^{-13}\cdot\frac{\langle K^{0}|\bar{d}_{L}\gamma_{\mu}s_{L}
 \bar{d}_{L}\gamma^{\mu}s_{L}|\overline{K^{0}}\rangle}{M_{K}} 
 \;\;\:{\rm (GeV^{-2})},
\end{eqnarray}
where $m_{c}$, $M_{W}$ and $M_{K}$ are the masses of the c quark, W boson
and the kaon respectively, and $G_{F}$ and $\alpha$ are the Fermi constant
and the fine structure constant.
$\theta_{W}$ and $\theta_{c}$ are the Weinberg angle and the Cabibbo angle
respectively.

\begin{figure}
 \leavevmode
 \epsfxsize=8cm
 \epsfysize=3cm
 \centerline{\epsfbox{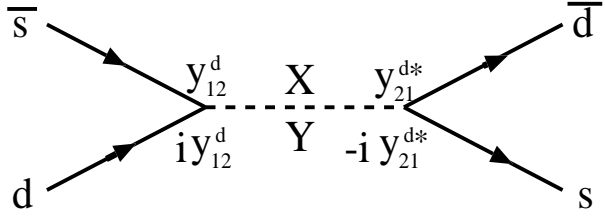}}
 \caption{Dominant contribution to ${\rm Im}M_{12}$.}
 \label{higgs-exchg}
\end{figure}

On the other hand, the dominant contribution to ${\rm Im}M_{12}$ comes from
the tree-level diagram shown by Fig.\ref{higgs-exchg},
because the box diagram mentioned above is real up to the phase of 
order $\varphi$ and cannot be the leading contribution.

This contribution is calculated by \cite{haber-nir}
\begin{eqnarray}
 M^{\mbox{\scriptsize tree}}_{12}&=&\frac{y^{d}_{12}y^{d\ast}_{21}}{M_{S}^{2}}
 \frac{\langle K^{0}|\bar{d}_{L}s_{R}\bar{d}_{R}s_{L}|\overline{K^{0}}\rangle}
 {M_{K}} \nonumber \\
 &\simeq& \left(\frac{v M_{\ast}}{M_{NP}^{2}}\right)^{2}
 \frac{\epsilon^{7}}{M_{S}^{2}}
 \frac{\langle K^{0}|\bar{d}_{L}s_{R}\bar{d}_{R}s_{L}|\overline{K^{0}}\rangle}
 {M_{K}},
\end{eqnarray}
where $M_{S}$ is the mass of the singlet Higgs field.

According to Ref.\cite{haber-nir},
\begin{equation}
 \frac{\langle K^{0}|\bar{d}_{L}s_{R}\bar{d}_{R}s_{L}|\overline{K^{0}}\rangle}
 {\langle K^{0}|\bar{d}_{L}\gamma_{\mu}s_{L}\bar{d}_{L}\gamma^{\mu}s_{L}|
 \overline{K^{0}}\rangle} \simeq 7.6,
\end{equation}
then 
\begin{equation}
 \left|\frac{M^{\mbox{\scriptsize tree}}_{12}}{M^{\mbox{\scriptsize box}}_{12}}
 \right| \simeq 10^{13}\: ({\rm GeV^{2}})\left(\frac{v M_{\ast}}
 {M_{NP}^{2}}\right)^{2}
 \frac{\epsilon^{7}}{M_{S}^{2}}\times 7.6 
 \simeq 2\times 10^{9}\left(\frac{v M_{\ast}}{M_{NP}^{2}}\right)^{2}
 \epsilon^{7}.
\end{equation}
Here we have assumed $M_{S}\simeq 200$ GeV.

Together with the fact that 
${\rm Arg}(M^{\mbox{\scriptsize tree}}_{12})=O(\varphi)$,
\begin{equation}
 |\epsilon_{K}|\simeq\frac{1}{2\sqrt{2}}\cdot\varphi\cdot 
 \left|\frac{M^{\mbox{\scriptsize tree}}_{12}}{M^{\mbox{\scriptsize box}}_{12}}
 \right|
 \simeq \frac{1}{\sqrt{2}}\cdot 10^{9}\cdot\epsilon^{8}\left(\frac{v}
 {M_{NP}}\right)^{2}
 \frac{w}{M_{NP}^{2}}
 \left(\frac{M_{\ast}}{M_{NP}}\right)^{3}
 \simeq 10^{-3}.
\end{equation}

\section{Neutrino} \label{neutrino}
In our scenario, the fundamental scale $M_{\ast}$ is of order TeV scale,
so at first sight it does not seem that the see-saw mechanism, 
which requires an intermediate scale around $10^{10}$ GeV, can be applied.
However, by using the volume factor suppression mentioned 
in Section.~\ref{ferm-hiechy} we can realize the small masses of the neutrinos.

We shall introduce a new extra dimension and denote its radius as $R_{2}$.
Let the right-handed neutrinos $\nu_{Ri}$ ($i=1,2,3$) feel this new 
extra dimension, so that neutrino Yukawa couplings $h^{\nu}_{ij}$ are
suppressed by a large volume factor $1/\sqrt{2\pi M_{\ast}R_{2}}$
and the Dirac masses of neutrinos can become small enough 
\cite{ark-dim-mch,dvl-smnv}.

Denote the Majorana mass scale of the right-handed neutrino as $M_{N}$,
then according to the see-saw mechanism left-handed neutrino mass matrix
$m_{\nu}$ is calculated as,
\begin{equation}
 m_{\nu}\simeq \frac{v^{2}}{M_{N}}\frac{1}{2\pi M_{\ast}R_{2}}
 \left(\begin{array}{ccc} \epsilon^{2} & \epsilon & \epsilon \\
 \epsilon & 1 & 1 \\ \epsilon & 1 & 1 \end{array}\right),
\end{equation}
where $v=174$ GeV.
This realizes the large mixing between $\nu_{\mu}$-$\nu_{\tau}$ expected from
the atmospheric neutrino and the small mixing between $\nu_{e}$-$\nu_{\mu}$ 
expected from the small angle MSW solution simultaneously \cite{yoshioka}.

For example, if we assume $M_{N}\simeq M_{\ast}$, we should set 
$R_{2}^{-1}\simeq 2$~keV in order to obtain 
$m_{\nu_{\tau}}\simeq 10^{-1}$~eV.

We shall list some notable comments.

First, if we try to suppress $m_{\nu}$ only by the volume factor
without the see-saw mechanism,
then $R_{2}$ becomes so large that the extra dimension will be observed by
the gravity experiment.
Thus we must either use the see-saw mechanism or let $\nu_{Ri}$ feel
more than one extra dimensions in order to realize 
the experimentally accepted small masses of the neutrinos
without contradiction to the gravity experiments.

Second, the effect of the running of $h^{\nu}_{ij}$ between  
$M_{NP}$ and $R_{2}^{-1}$ is not expected to be very large
by the same reason mentioned at the footnote in Section.\ref{ferm-hiechy}.
Then we have neglected this running effect in the above discussion.

Finally, it is worthy to note that the infinite Kaluza-Klein modes of 
$\nu_{Ri}$ can correspond to the sterile neutrinos from the phenomenological
point of view \cite{dvl-smnv}.

\section{Strong CP problem} \label{strong-cp}
The axion scenario is the most convincing solution to the strong CP problem.
Similarly to the previous section, however, the axion scenario also needs 
the intermediate scale that the Peccei-Quinn symmetry is broken at.
Here we shall avoid this difficulty by using the power-law running of
coupling constants, which is characteristic of the context of 
large extra dimensions.

Assume that the axion, which is confined to our four dimensional wall, 
interacts with spinor fields $\psi$ and $\bar{\psi}$, which feel 
two extra dimensions whose radii are both $R_{3}$.
\begin{displaymath}
 {\cal L}_{a\psi\bar{\psi}}=g_{\psi}a\psi\bar{\psi},
\end{displaymath}
where $a$ represents the axion field.

In this case a wavefunction renormalization factor of the axion $Z_{a}$
will scale according to the power-law running above the scale 
$\mu_{3}\equiv R_{3}^{-1}$.
\begin{equation}
 Z_{a}=1-cg_{\psi}^{2}\left(\frac{\Lambda}{\mu_{3}}\right)^{4}+\cdots,
\end{equation}
where $a(\Lambda)=Z_{a}^{1/2}a(\mu_{3})$, $\Lambda$ is a cut-off and
$c$ is an $O(1)$ constant.

\begin{figure}
\leavevmode
\epsfxsize=8cm
\epsfysize=3cm
\centerline{\epsfbox{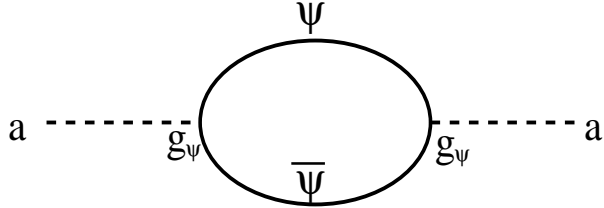}}
\caption{Relevant diagram to $Z_{a}$.}
\end{figure}

\begin{equation}
 g_{\psi}(\Lambda)=Z_{a\psi\bar{\psi}}Z_{a}^{-1/2}Z_{\psi}^{-1/2}
 Z_{\bar{\psi}}^{-1/2}g_{\psi}(\mu_{3}),
\end{equation}
where $Z_{a\psi\bar{\psi}}$ is a vertex renormalization factor,
$Z_{\psi}$ and $Z_{\bar{\psi}}$ are wavefunction renormalization
factors of $\psi$ and $\bar{\psi}$ respectively.

Here if we assume that the power-law running of $Z_{a}$ is the strongest
of the runnings of the $Z$-factors, then
\begin{equation}
 g_{\psi}^{-2}(\Lambda)\simeq\left\{1-cg_{\psi}^{2}\left(
 \frac{\Lambda}{\mu_{3}}\right)^{4}\right\}g_{\psi}^{-2}(\mu_{3}).
\end{equation}

Next we put another assumption that the value of $g_{\psi}$ 
at the Peccei-Quinn symmetry breaking scale (PQ scale) $M_{PQ}$ 
is much larger than $g_{\psi}(\mu_{3})$.
Then we obtain 
\begin{equation}
 g_{\psi}(\mu_{3})\simeq \left(\frac{\mu_{3}}{M_{PQ}}\right)^{2}.
\end{equation}

From the assumption, this suppression mainly comes from the power-law running
of $Z_{a}$, so that all couplings including the axion field $a$ are expected
to receive the same suppression factor as that of $g_{\psi}$.
For example, the coupling $-(1/64\pi^{2})(a/M_{PQ})\epsilon_{\mu\nu\rho\sigma}
F_{\alpha}^{\mu\nu}F_{\alpha}^{\rho\sigma}$ receives the suppression factor
$(\mu_{3}/M_{PQ})^{2}$ below the scale $\mu_{3}$, thus effective PQ scale
$f_{PQ}$ becomes
\begin{equation}
 f_{PQ}\simeq\left(\frac{M_{PQ}}{\mu_{3}}\right)^{2}M_{PQ}.
\end{equation}

For instance, if we set $M_{PQ}\simeq M_{\ast}$ and $\mu_{3}=10$~GeV, then
we shall obtain $f_{PQ}\simeq 10^{10}$~GeV and this value satisfies 
the cosmological constraint.\footnote{In Ref.\cite{chang} 
the axion field itself is suppose to be 
the bulk field and the constraint:
$10^{9}\mbox{ GeV}<f_{PQ}<10^{15}\mbox{ GeV}$ is satisfied 
by the volume factor suppression.}

\section{Conclusions} \label{conclusions}
We showed in the context of the large extra dimensions enough CP violation 
can be obtained from the spontaneous breakdown in a simple non-SUSY model, 
which is usually considered not
to cause the spontaneous CP violation and estimated $\epsilon_{K}$
to be of order $10^{-3}$ consistent with the experimental value.
It is appealing that the same volume factors are used
to generate both adequate smallness of the CP violation 
and the hierarchy among Yukawa couplings,
which is used to suppress the flavor changing neutral current (FCNC).

Our scenario does not depend on the Higgs potential that has 
a CP violating minimum and no extra symmetries are introduced, 
so we can easily generalize our model to models with 
more complicated Higgs sector.
For example, two-Higgs-doublet standard model with the exact discrete symmetry
of natural flavor conservation can also cause the SCPV in our scenario.
The essence of our scenario is the existence of the suppressed 
extra Yukawa matrices that have complex phases and 
become main sources of the CP violation.
Another work in this direction is, for example Ref.\cite{masip-rasin},
in which extra Higgs doublets are introduced and Peccei-Quinn-like 
approximate symmetry are used to suppress the dangerous FCNC and 
the CP violation to the observed level.
On the other hand, we have used the volume factor suppression in the context
of large extra dimensions instead of some approximate symmetries.

In our scenario, naive see-saw mechanism or axion scenario, which need
an intermediate scale around $10^{10}$~GeV, cannot be applied
since the fundamental scale $M_{\ast}$ is TeV scale.
However, these difficulties can be avoided by making use of 
the volume factor suppression or the power-law running of couplings, 
which are characteristic of the context of large extra dimensions.

One can also consider the scenario that the Yukawa couplings are generated 
at the scale
$M_{\ast}$ and realize the hierarchy among the Yukawa couplings as
the hierarchy among quasi infrared fixed points (QFPs) \cite{abel-king},
but in such a case the coupling constants $\hat{y}^{x}_{ij}$ would 
become too small to realize the realistic value of $\epsilon_{K}$.
Also, in this case $M_{NP}$ has to be lift up to 
$M_{\ast}$, we would need some mechanism that suppresses the power-law running
of the gauge coupling constants, for example N=4 SUSY.

So far we have not assumed supersymmetry because SUSY models have 
so many sources of the CP violation that the observed CP violation 
can be obtained without our scenario.
However our scenario can be applied
in SUSY models if we adopt an appropriate SUSY breaking mechanism 
(for example Scherk-Schwarz mechanism \cite{scherk,anton-dimo}) in which 
super-particles are heavy enough, squark masses are degenerate and so on.
In this case, we can say that our scenario extends the parameter space
that gives the observed CP violation in the SCPV.

We collect the example values of all scales used here,
\begin{eqnarray*}
 &&R_{2}^{-1}\simeq 2\mbox{ keV},\;\;R_{3}^{-1}\simeq 10\mbox{ GeV},\;\;
 \langle S\rangle\simeq 200\mbox{ GeV},\;\;R_{1}^{-1}\simeq 300\mbox{ GeV}, \\
 &&M_{NP}\simeq 3\mbox{ TeV},\;\;
 M_{N}\simeq M_{PQ}\simeq M_{\ast}\simeq 10\mbox{ TeV}.
\end{eqnarray*}
Of course, there are various other possibilities for their values.

Here we introduced new five extra dimensions.
These must satisfy the relation Eq.(\ref{rlt-pf}).
In the case of $n=6$, which is motivated by the superstring theory,
the remaining radius is about $(4\mbox{ MeV})^{-1}$ and satisfies
the constraint from the gravity experiment.

If we suppose our scenario to be right, we can represents each scale as 
a function of $R_{1}^{-1}$.
\begin{eqnarray}
 &&M_{\ast}\simeq 36R_{1}^{-1},\;\; M_{NP}\simeq 1.8\times 10^{2}R_{1}^{-1/2},
 \;\; R_{2}^{-1}\simeq 2.7\times 10^{-11}R_{1}^{-2}, \nonumber \\
 &&R_{3}^{-1}\simeq 2.1\times 10^{-3}R_{1}^{-3/2},\;\; 
 R_{4}\simeq 1.4\times 10^{-5}R_{1}^{-1},
\end{eqnarray}
where the unit is GeV, and $R_{4}$ is the remaining radius in the case of
$n=6$.

Of course, the hierarchy among the radii of the extra dimensions,
which is assumed here, must be explained for completeness.

\vspace{1cm}
\mbox{}\linebreak
{\Large\bf Acknowledgements}

\mbox{}\linebreak
The author would like to thank N.Sakai for useful advice and
N.Maru and T.Matsuda for valuable discussion and information.

\end{document}